\let\oldvec\vec
\documentclass[runningheads]{llncs}
\let\vec\oldvec

\usepackage{times}
\usepackage{amssymb}
\usepackage{amsmath}
\usepackage{graphicx}

\newtheorem{mytheorem}{Theorem}
\newtheorem{mydefinition}{Definition}
\newtheorem{myexample}{Example}

\newcommand{\myqed}{\mbox{$\Box$}}

\begin{document}

\title{Group Envy Freeness and Group Pareto Efficiency \\ in Fair Division with Indivisible Items}
\titlerunning{Group Envy Freeness and Group Pareto Efficiency in Fair Division}

\author{Martin Aleksandrov \and Toby Walsh}
\authorrunning{M. Aleksandrov \and T. Walsh}

\institute{Technical University of Berlin, Berlin, Germany\\ \email{\{martin.aleksandrov,toby.walsh\}@tu-berlin.de}}

\maketitle              

\begin{abstract}
We study the fair division of items to agents supposing that agents can form groups. We thus give natural generalizations of popular concepts such as envy-freeness and Pareto efficiency to groups of fixed sizes. \emph{Group envy-freeness} requires that no group envies another group. \emph{Group Pareto efficiency} requires that no group can be made better off without another group be made worse off. We study these new group properties from an axiomatic viewpoint. We thus propose new fairness taxonomies that generalize existing taxonomies. We further study \emph{near} versions of these group properties as allocations for some of them may not exist. We finally give three prices of \emph{group} fairness between group properties for three common social welfares (i.e.\ utilitarian, egalitarian and Nash).

\end{abstract}

\section{Introduction}

Fair divisions become more and more challenging in the present world due to the ever-increasing demand for resources. This pressure forces us to achieve more complex allocations with less available resources. An especially challenging case of fair division deals with the allocation of \emph{free-of-charge} and \emph{indivisible} items (i.e.\ items cannot be divided, items cannot be purchased) to agents cooperating in \emph{groups} (i.e.\ each agent maximizes multiple objectives) in the absence of information about these groups and their group preferences. For example, food banks in Australia give away perishable food products to charities that feed different \emph{groups} of the community (e.g. Muslims) \cite{pia2014,pia2016}. As a second example, social services in Germany provide medical benefits, donated food and affordable education to thousands of refugees and their \emph{families}. We often do not know the group members or how they share group preferences for resources. Some other examples are the allocations of office rooms to research groups \cite{bouveret2017}, cake to groups of guests \cite{brams1996,halevi2015}, land to families \cite{kokoye2013}, hospital rooms to medical teams \cite{smet2016} and memory to computer networks \cite{parkes2015}. 

In this paper, we consider the fair division of items to agents under several assumptions. For example, the collection of items can be a mixture of goods and bads (e.g.\ meals, chores) \cite{aziz2017,bogomolnaia2016,lumet2012}. We thus assume that each agent has some aggregate utility for a given bundle of items of another agent. However, these utilities can be shared arbitrarily among the sub-bundles of the bundle (e.g. monotonically, additively, modularly, etc.). As another example, the agents can form groups in an arbitrarily manner. We thus assume that each group has some aggregate utility for a given bundle of items of another group. As in \cite{halevi2015}, we consider arithmetic-mean group utilities. We study this problem for five main reasons. First, people form groups naturally in practice (e.g.\ families, teams, countries). Second, group preferences are more expressive than individual preferences but also more complex (e.g.\ complementarities, substitutabilities). Third, we seek new group properties as many existing ones may be too demanding (e.g.\ coalitional fairness). Fourth, the principles in which groups form are normally not known. Fifth, with arithmetic-mean group utilities, we generalize existing fairness taxonomies \cite{aziz2018,aziz2015epera} and characterization results for Pareto efficiency \cite{bogomolnaia2001}. 

Two of the most important criteria in fair division are envy-freeness (i.e.\ no agent envies another agent) and Pareto efficiency (i.e.\ no agent can be made better off without another agent be made worse off) \cite{brams2000,brams2005,clippel2008,weller1985}. We propose \emph{new} generalizations of these concepts for groups of fixed sizes. \emph{Group envy-freeness} requires that no group envies another group. \emph{Group Pareto efficiency} requires that no group can be made better off without another group be made worse off. We thus construct new sets of fairness properties, that let us interpolate between envy-freeness and proportionality (i.e.\ each agent gets $1/n$ their total utility for bundles), and utilitarian efficiency (i.e.\ the sum of agent's utilities is maximized) and Pareto efficiency. There is a reason why we focus on these two common properties and say not on other attractive properties such as group strategy-proofness. Group strategy-proofness may not be achievable with limited knowledge of the groups \cite{aleksandrov2017ijcai}. By comparison, both group envy-freeness and group Pareto efficiency are achievable. For example, the allocation of each bundle uniformly at random among agents is group envy-free, and the allocation of each bundle to a given agent is group Pareto efficient. This example further motivates why we study these two properties in isolation. In some instances, no allocation satisfies them in combination. 

Common computational problems about group envy-freeness and group Pareto efficiency are inherently intractable even for problems of relatively small sizes \cite{bliem2016,bouveret2008,keijzer2009}. For this reason, we focus on the axiomatic analysis of these properties. We propose a taxonomy of $n$ layers of group envy-freeness properties such that group envy-freeness at layer $k$ implies (in a logical sense) group envy-freeness at layer $k+1$. This is perhaps a good news because envy-free allocations often do \emph{not} exist and, as we show, allocations satisfying some properties in our taxonomy \emph{always} exist. We propose another taxonomy of $n$ layers of group Pareto efficiency properties such that group Pareto efficiency at layer $k+1$ implies group Pareto efficiency at layer $k$. Nevertheless, it is not harder to achieve group Pareto efficiency than Pareto efficiency and such allocations still \emph{always} exists. We also consider $\alpha$-taxonomies of \emph{near} group envy-freeness and \emph{near} group Pareto efficiency properties for each $\alpha\in[0,1]$. We finally use prices of \emph{group} fairness to measure the ``loss'' in welfare efficiency between group properties. 

Our paper is organized as follows. We next discuss related work and define our notions. We then present our taxonomy for group envy-freeness in the cases in which agents might be envy of groups (Theorem~\ref{thm:one}), groups might be envy of agents (Theorem~\ref{thm:two}) and groups might be envy of groups (Theorem~\ref{thm:three}). We continue with our taxonomy for group Pareto efficiency (Theorem~\ref{thm:four}) and generalize an important result from Pareto efficiency to group Pareto efficiency (Theorem~\ref{thm:five}). Further, we propose taxonomies of properties approximating group envy-freeness and group Pareto efficiency. Finally, we give the prices of group fairness (Theorem~\ref{thm:six}) and conclude our work.

\section{Related work}\label{rel}

Group fairness has been studied in the literature. Some notions compare the bundle of each group of agents to the bundle of any other group of agents based on Pareto dominance (i.e.\ all agents are weakly happier, and some agents are strictly happier) preference relations (e.g.\ coalitional fairness, strict fairness) \cite{debreu1967,husseinov2011,lahaie2009,schmeidler1972,varian1974,vind1964,zhou1992}. 
Coalitional fairness implies both envy-freeness and Pareto efficiency. Perhaps this might be too demanding in practice as very often such allocations do not exist. For example, for a given allocation, it requires complete knowledge of agents' utilities for any bundles of items of any size in the allocation, whereas our notions require \emph{only} knowledge of agents' utilities for their own bundles and the bundles of other agents in the allocation. Other group fairness notions are based on the idea that the bundle of each group should be perceived as fair by as many agents in the group as possible (e.g.\ unanimously envy-freeness, $h$-democratic fairness, majority envy-freeness) \cite{halevi2018,suksompong2018}. The authors suppose that the groups are disjoint and known (e.g.\ families), and the utilities of agents for items are known, whereas we suppose that the groups are \emph{unknown}, thus possibly \emph{overlap}, and the utilities of agents are in a \emph{bundle} form.  

More group fairness notions have been studied in the context of cake-cutting (e.g.\  arithmetic-mean-proportionality, geometric-mean-proportionality, minimum-proporti-onality, median-proportionality) \cite{halevi2015}. These notions compare the aggregate bundle of each group of agents to their proportional (wrt the number of groups) aggregate bundle of all items. \emph{Unlike} us, the authors assume that the group members and their monotonic valuations are part of the common knowledge. Group envy-freeness notions are also already used in combinatorial auctions with additive quasi-linear utilities and monetary transfers (e.g.\ envy-freeness of an individual towards a group, envy-freeness of a group towards a group) \cite{todo2011}. The authors assume that the agents' utilities for items and item prices are known. Conceptually, our notions of group envy-freeness resemble these notions but they do \emph{not} use prices. We additionally study notions of near group fairness. Our near group fairness notions for groups of agents are inspired by $\alpha$-fairness for individual agents \cite{borsuk1933,dubins1961,hill1983,steinhaus1948,stone1942}.

Most of these existing works consider allocating divisible resources (e.g.\ land, cake) with money (e.g.\ exchange economies), whereas we consider allocating \emph{indivisible} items \emph{without} money. We further \emph{cannot} directly apply most of these existing properties to our setting with unknown groups, bundle utilities and priceless items. As a result, we \emph{cannot} directly inherit any of the existing results. In contrast, we can apply our group properties in settings in which the group members and their preferences are actually known. Therefore, our results are valid in some existing settings. Our properties are \emph{new} and cannot be defined using the existing fairness framework proposed in \cite{aziz2018}. Moreover, existing works are somehow related to our properties of group envy-freeness. However, we additionally propose properties of group Pareto efficiency. Also, most existing properties may \emph{not} be guaranteed even with a single indivisible item (e.g.\ coalitional fairness). By comparison, \emph{many} of our group envy-freeness properties and \emph{all} of our group Pareto efficiency properties can be guaranteed. Furthermore, we use \emph{new} prices of fairness for our group properties similarly as for other properties in other settings \cite{aleksandrov2017mcm,bertsimas2011,kaleta2014,nicosia2017}. Finally, several related models are studied in \cite{manurangsi2017,suksompong2016,yokoo2003}. However, none of these focuses on axiomatic properties such as ours.

\section{Preliminaries}\label{sec:pre}

We consider a set $N=\lbrace a_1,\ldots,a_n\rbrace$ of agents and a set $O=\lbrace o_1,\ldots,o_m\rbrace$ of indivisible items. We write $\pi=(\pi_1,\ldots,\pi_n)$ for an \emph{allocation} of the items from $O$ to the agents from $N$ with (1) $\cup_{a\in N}^n\pi_a=O$ and (2) $\forall a,b\in N,a\not=b:\pi_a\cap\pi_b=\emptyset$, where $\pi_a,\pi_b$ denote the bundles of items of agents $a,b\in N$ in $\pi$. We suppose that agents form groups. We thus write $\pi_G$ for the bundle $\cup_{a\in G}\pi_a$ of items of group $G$, and $u_G(\pi_{H})$ for the \emph{utility} of $G$ for the bundle $\pi_{H}$ of items of group $H$. We assume \emph{arithmetic-mean} group utilities. That is, $u_{G}(\pi_{G})=\frac{1}{k}\cdot\sum_{a\in G} u_{a}(\pi_a)$ and $u_{G}(\pi_{H})=\frac{1}{k\cdot h}\cdot\sum_{a\in G}\sum_{b\in H} u_{a}(\pi_b)$, where the group $G$ has $k$ agents, the group $H$ has $h$ agents and the utility $u_a(\pi_b)\in\mathbb{R}^{\geq 0}$ can be arbitrary for any agents $a,b\in N$ (i.e.\ monotonic, additive, modular, etc.).

We next define our group fairness properties. Group envy-freeness captures the envy of a group towards another group. Group Pareto efficiency captures the fact that we cannot make each group weakly better off, and some group strictly better off. These properties strictly generalize envy-freeness and Pareto efficiency whenever the group sizes are fixed. Near group fairness is a relaxation of group fairness. 

\begin{mydefinition} \emph{\bf (group envy-freeness)} For $k,h\in\lbrace 1,\ldots,n\rbrace$,
an allocation $\pi$ is \emph{$(k,h)$-group envy-free (or simply GEF$_{k,h}$)} iff, for each group $G$ of $k$ agents and each group $H$ of $h$ agents, $u_{G}(\pi_{G})\geq u_{G}(\pi_{H})$ holds. 
\end{mydefinition}

\begin{mydefinition} \emph{\bf (group Pareto efficiency)} For $k\in\lbrace 1,\ldots,n\rbrace$, an allocation $\pi$ is \emph{$k$-group Pareto efficient (or simply GPE$_k$)} iff, there is no other allocation $\pi^{\prime}$ such that $u_{G}(\pi^{\prime}_{G})\geq u_{G}(\pi_{G})$ holds for each group $G$ of $k$ agents, and $u_{H}(\pi^{\prime}_{H})> u_{H}(\pi_{H})$ holds for some group $H$ of $k$ agents.
\end{mydefinition}

\begin{mydefinition} \emph{\bf (near group envy-freeness)} For $k,h\in\lbrace 1,\ldots,n\rbrace$ and $\alpha\in\mathbb{R}^{[0,1]}$, an allocation $\pi$ is \emph{near $(k,h)$-group envy-free wrt $\alpha$ (or simply GEF$^{\alpha}_{k,h}$)} iff, for each group $G$ of $k$ agents and each group $H$ of $h$ agents, $u_{G}(\pi_{G})\geq \alpha\cdot u_{G}(\pi_{H})$ holds. 
\end{mydefinition}

\begin{mydefinition} \emph{\bf (near group Pareto efficiency)} For $k\in\lbrace 1,\ldots,n\rbrace$ and $\alpha\in\mathbb{R}^{[0,1]}$, an allocation $\pi$ is \emph{near $k$-group Pareto efficient wrt $\alpha$ (or simply GPE$^{\alpha}_k$)} iff, there is no other allocation $\pi^{\prime}$ such that $\alpha\cdot u_{G}(\pi^{\prime}_{G})\geq u_{G}(\pi_{G})$ holds for each group $G$ of $k$ agents, and $\alpha\cdot u_{H}(\pi^{\prime}_{H})> u_{H}(\pi_{H})$ holds for some group $H$ of $k$ agents.
\end{mydefinition}

We use prices to measure the ``loss'' in the welfare $w(\pi)$ between these properties in a given allocation $\pi$. The \emph{price of group envy-freeness} $p^w_{\text{\tiny{GEF}}}$ is \text{\footnotesize{$\max_{k,h}\frac{\max_{\pi_1} w(\pi_1)}{\min_{\pi_2} w(\pi_2)}$}} where $\pi_1$ is a $(h,h)$-group envy-free and $\pi_2$ is a $(k,k)$-group envy-free with $h\leq k$. The \emph{price of group Pareto efficiency} $p^w_{\text{\tiny{GPE}}}$ is \text{\footnotesize{$\max_{k,h}\frac{\max_{\pi_1} w(\pi_1)}{\min_{\pi_2} w(\pi_2)}$}} where $\pi_1$ is a $h$-group Pareto efficient and $\pi_2$ is a $k$-group Pareto efficient with $h\geq k$. The \emph{price of group fairness} $p^w_{\text{\tiny{FAIR}}}$ is \text{\footnotesize{$\max_{k}\frac{\max_{\pi_1} w(\pi_1)}{\min_{\pi_2} w(\pi_2)}$}} where $\pi_1$ is a $(k,k)$-group envy-free and $\pi_2$ is a $k$-group Pareto efficient. We consider these prices for common welfares such as the utilitarian welfare $u(\pi)=\sum_{a\in N} u_{a}(\pi_{a})$, the egalitarian welfare $e(\pi)=\min_{a\in N} u_{a}(\pi_{a})$ and the Nash welfare $n(\pi)=\prod_{a\in N} u_{a}(\pi_{a})$.

Finally, we write $\Pi_H$ for the \emph{expected allocation} of group $H$ that assigns a probability value to each bundle of items, and $\overline{u}_G(\Pi_{H})$ for the \emph{expected utility} of group $G$ for $\Pi_H$. We observe that we can define our group properties in terms of expected utilities of groups for expected allocations of groups.

\section{Group envy freeness}\label{sec:gef}

We start with group envy-freeness for arithmetic-mean group utilities. Our first main result is to give a taxonomy of strict implications between group envy-freeness notions for groups of fixed sizes (i.e.\ GEF$_{k,h}$ for fixed $k,h\in[1,n)$). We present the taxonomy in Figure~\ref{fig:eftax}.

\begin{figure}[ht]
\begin{center}
\begin{tabular}{ccc}
GEF$_{k,h}$ & $\Rightarrow$ & GEF$_{k,h+1}$ \\
$\Downarrow$ &  & $\Downarrow$ \\
GEF$_{k+1,h}$ & $\Rightarrow$ & GEF$_{k+1,h+1}$ \\
\end{tabular}
\end{center}
\caption{A taxonomy of group envy-freeness properties for fixed $k,h\in[1,n)$.}
\label{fig:eftax}
\end{figure}

Our taxonomy contains $n^2$ group envy-freeness axiomatic properties. By definition, we observe that $(1,1)$-group envy-freeness is equivalent to envy-freeness (or simply EF) and $(1,n)$-group envy-freeness is equivalent to proportionality (or simply PROP). Moreover, we observe that $(n,1)$-group envy-freeness captures the envy of the group of all agents towards each agent. We call this property \emph{grand envy-freeness} (or simply gEF). $(n,n)$-group envy-freeness is trivially satisfied by any allocation. In our taxonomy, we can interpolate between envy-freeness and proportionality, and even beyond. From this perspective, our taxonomy generalizes existing taxonomies of fairness concepts for individual agents with additive utilities \cite{aziz2018,aziz2015epera}. We next prove the implications in our taxonomy. For this purpose, we distinguish between \emph{agent-group} properties (i.e.\ $(1,h)$-group envy-freeness), \emph{group-agent} properties (i.e.\ $(k,1)$-group envy-freeness) and \emph{group-group} properties (i.e.\ $(k,h)$-group envy-freeness) for $k\in[1,n]$ and $h\in[1,n]$.

\subsubsection{Agent-group envy-freeness}

We now consider $n$ properties for agent-group envy-free-ness of actual allocations that capture the envy an individual agent might have towards a group of other agents. These properties let us move from envy-freeness to proportionality (i.e.\ there is $h\in[1,n]$ such that ``EF $\Rightarrow$ GEF$_{1,h}$ $\Rightarrow$ PROP''). If an agent is envy-free of a group of $h\in[1,n]$ agents, then they are envy-free of a group of $q\geq h$ agents.

\begin{mytheorem}\label{thm:one}
For $h\in[1,n]$, $q\in[h,n]$ and arithmetic-mean group utilities, we have that \emph{GEF$_{1,h}$} implies \emph{GEF$_{1,q}$} holds.
\end{mytheorem}

\begin{myproof}
Let us pick an allocation $\pi$. We show the result by induction on $i\in[h,q]$. In the base case, let $i$ be equal to $h$. The result follows trivially in this case. In the induction hypothesis, suppose that $\pi$ is $(1,i)$-group envy-free for $i<q$. In the step case, let $i$ be equal to $q$. By the hypothesis, we know that $\pi$ is $(1,q-1)$-group envy-free. For the sake of contradiction, let us suppose that $\pi$ is not $(1,q)$-group envy-free. Consequently, there is a group of $q$ agents and an agent, say $G=\lbrace a_{1},\ldots,a_{q}\rbrace$ and $a\not\in G$, such that inequality~(\ref{eq:agefone}) holds for $G$ and $a$, and inequality~(\ref{eq:ageftwo}) holds for $G$, $a$ and each agent $a_j\in G$.

\begin{equation}\label{eq:agefone}
{u}_{a}(\pi_a)<u_a(\pi_G)=\frac{1}{q}\cdot\sum_{b\in G} {u}_{a}(\pi_b)
\end{equation} 
\begin{equation}\label{eq:ageftwo}
{u}_{a}(\pi_a)\geq u_a(\pi_{G\setminus\lbrace a_j\rbrace})=\frac{1}{(q-1)}\cdot\sum_{b\in G\setminus\lbrace a_j\rbrace} {u}_{a}(\pi_b)  
\end{equation} 

We derive ${u}_a(\pi_a)<{u}_a(\pi_{a_j})$ for each $a_j\in G$. Let us now form a group of $(q-1)$ agents from $G$, say $G\setminus\lbrace a_{q}\rbrace$. Agent $a$ assigns arithmetic-mean value to the allocation of this group that is larger than the value they assign to their own allocation. This contradicts with the induction hypothesis. Hence, $\pi$ is $(1,q)$-group envy-free. The result follows. \myqed
\end{myproof}

By Theorem~\ref{thm:one}, we conclude that $(1,h)$-group envy-freeness implies $(1,h+1)$-group envy-freeness for $h\in[1,n)$. The opposite direction does not hold. Indeed, $(1,q)$-group envy-freeness is a weaker property than $(1,h)$-group envy-freeness for $q>h$. We illustrate this in Example~\ref{exp:one}. 

\begin{myexample}\label{exp:one}
Let us consider the fair division of 3 items $o_1,o_2,o_3$ between 3 agents $a_1,a_2,a_3$. Further, let the utilities of agent $a_1$ for the items be $1$, $3/2$ and $2$, those of agent $a_2$ be $3/2$, $2$, and $1$, and the ones of agent $a_3$ be $2$, $1$ and $3/2$ respectively. Now, consider the allocation $\pi$ that gives $o_2$ to $a_1$, $o_1$ to $a_2$ and $o_3$ to $a_3$. Each agent receives in $\pi$ utility $3/2$. Hence, this allocation is not $(1,1)$-group envy-free (i.e.\ envy-free) as each agent assigns in it utility $2$ to one of the other agents. In contrast, they assign in $\pi$ utility $3/2$ to the group of all agents. We conclude that $\pi$ is $(1,3)$-group envy-free (i.e.\ proportional). \myqed
\end{myexample}

The result in Example~\ref{exp:one} crucially depends on the fact that there are 3 agents in the problem. With 2 agents, agent-group envy-freeness is equivalent to envy-freeness which itself is equivalent to proportionality. Finally, Theorem~\ref{thm:one} and Example~\ref{exp:one} hold for expected allocations as well.

\subsubsection{Group-agent envy-freeness}

We next consider $n$ properties for group-agent envy-freeness of actual allocations that capture the envy a group of agents might have towards an individual agent outside the group. These properties let us move from envy-freeness to grand envy-freeness (i.e.\ there is $k\in[1,n]$ such that ``EF $\Rightarrow$ GEF$_{k,1}$ $\Rightarrow$ gEF''). If a group of $k\in[1,n]$ agents is envy-free of a given agent, then a group of $p\geq k$ agents is envy-free of this agent.

\begin{mytheorem}\label{thm:two}
For $k\in[1,n]$, $p\in[k,n]$ and arithmetic-mean group utilities, we have that \emph{GEF$_{k,1}$} implies \emph{GEF$_{p,1}$} holds.
\end{mytheorem}

\begin{myproof}
Let us pick an allocation $\pi$. As in the proof of Theorem~\ref{thm:one}, we show the result by induction on $i\in[k,p]$. The most interesting case is the step case. Let $i$ be equal to $p$ and suppose that $\pi$ is $(p-1,1)$-group envy-free. For the sake of contradiction, let us suppose that $\pi$ is not $(p,1)$-group envy-free. Consequently, there is a group of $p$ agents and an agent, say $G=\lbrace a_{1},\ldots,a_{p}\rbrace$ and $a\not\in G$, such that inequality~(\ref{eq:agefthree}) holds for $G$ and $a$, and inequality~(\ref{eq:ageffour}) holds for $G$, $a$ and each $a_j\in G$.

\begin{equation}\label{eq:agefthree}
p\cdot u_G(\pi_G)=\sum_{b\in G} {u}_b(\pi_b)<\sum_{b\in G} {u}_b(\pi_a)
\end{equation} 
\begin{equation}\label{eq:ageffour}
\begin{split}
(p-1)\cdot u_{G\setminus\lbrace a_j\rbrace}(\pi_{G\setminus\lbrace a_j\rbrace})=\sum_{b\in G\setminus\lbrace a_j\rbrace} {u}_b(\pi_b)\geq \sum_{b\in G\setminus\lbrace a_j\rbrace} {u}_b(\pi_a) 
\end{split} 
\end{equation} 

We derive ${u}_{a_j}(\pi_{a_j})<{u}_{a_j}(\pi_{a})$ for each $a_j\in G$. Let us now form a group of $(p-1)$ agents from $G$, say $G\setminus\lbrace a_{p}\rbrace$. This group assigns arithmetic-mean value to the allocation of agent $a$ that is larger than the arithmetic-mean value they assign to their own allocation. This contradicts with the fact that $\pi$ is $(p-1,1)$-group envy-free. We therefore conclude that $\pi$ is $(p,1)$-group envy-free.\myqed
\end{myproof}

By Theorem~\ref{thm:two}, we conclude that $(k,1)$-group envy-freeness implies $(k+1,1)$-group envy-freeness for $k\in[1,n)$. However, $(p,1)$-group envy-freeness is a weaker property than $(k,1)$-group envy-freeness for $p>k$. We illustrate this in Example~\ref{exp:two}. 

\begin{myexample}\label{exp:two}
Let us consider again the instance in Example~\ref{exp:one} and the allocation $\pi$ that gives to each agent the item they value with $3/2$. We confirmed that $\pi$ is not $(1,1)$-group envy-free (i.e.\ envy-free). However, $\pi$ is $(3,1)$-group envy-free (i.e.\ grand envy-free) because the group of all agents assigns in $\pi$ utility $3/2$ to their own allocation and utility $3/2$ to the allocation of each other agent.\myqed
\end{myexample}

The choice of 3 agents in the problem in Example~\ref{exp:two} is again crucial. With 2 agents, group-agent envy-freeness is equivalent to envy-freeness and proportionality. Finally, Theorem~\ref{thm:two} and Example~\ref{exp:two} hold for expected allocations as well.

\subsubsection{Group-group envy-freeness}

We finally consider $n^2$ properties for group-group envy-freeness of actual allocations that captures the envy of a group of $k$ agents towards another group of $h$ agents. Similarly, we prove a number of implications between such properties for fixed parameters $k$, $h$ and $p\geq k$, $q\geq h$.

\begin{mytheorem}\label{thm:three}
For $k\in[1,n]$, $p\in[k,n]$, $h\in[1,n]$, $q\in[h,n]$ and arithmetic-mean group utilities, we have that \emph{GEF$_{k,h}$} implies \emph{GEF$_{p,q}$} holds.
\end{mytheorem}

\begin{myproof}
We prove by inductions that (1) $(p,h)$-group envy-freeness implies $(p,q)$-group envy-freeness for any $p\in[1,n]$, and that (2) $(k,h)$-group envy-freeness implies $(p,h)$-group-envy freeness for any $h\in[1,n]$. We can then immediately conclude the result. For $p=1$ in (1) and $h=1$ in (2), the base cases of the inductions follow from Theorems~\ref{thm:one} and~\ref{thm:two}. We start with (1). We consider only the step case. That is, let $\pi$ be an allocation that is $(p,q-1)$-group envy-free but not $(p,q)$-group envy-free. Hence, there are groups $G=\lbrace a_{1},\ldots,a_p\rbrace$ and $H=\lbrace b_1,\ldots,b_q\rbrace$ such that inequality (\ref{eq:ageffive}) holds for $G$ and $H$, and inequality (\ref{eq:agefsix}) holds for $G$, $H$ and each $b_j\in H$.

\begin{equation}\label{eq:ageffive}
\sum_{a\in G} u_a(\pi_a)<\frac{1}{q}\cdot\sum_{a\in G} \sum_{b\in H} u_a(\pi_b)
\end{equation} 
\begin{equation}\label{eq:agefsix}
\sum_{a\in G} u_a(\pi_a)\geq\frac{1}{(q-1)}\cdot\sum_{a\in G} \sum_{b\in H\setminus\lbrace b_j\rbrace} u_a(\pi_b)
\end{equation} 

We derive $\sum_{a\in G} u_a(\pi_a) <\sum_{a\in G} u_a(\pi_{b_j})$ for each $b_j\in H$ which leads to a contradiction with the $(p,q-1)$-group envy-freeness of $\pi$. We next prove (2) for $h=q$ in a similar fashion. Again, we consider only the step case. That is, let $\pi$ be an allocation that is $(p-1,q)$-group envy-free but not $(p,q)$-group envy-free. Hence, there are groups $G=\lbrace a_{1},\ldots,a_p\rbrace$ and $H=\lbrace b_1,\ldots,b_q\rbrace$ such that inequality (\ref{eq:ageffive}) holds for $G$ and $H$, and inequality (\ref{eq:agefseven}) holds for $G$, $H$ and each $a_j\in G$.

\begin{equation}\label{eq:agefseven}
\sum_{a\in G\setminus\lbrace a_j\rbrace} u_a(\pi_a)\geq\frac{1}{q}\cdot\sum_{a\in G\setminus\lbrace a_j\rbrace} \sum_{b\in H} u_a(\pi_b)
\end{equation} 

We obtain that $q\cdot u_{a_j}(\pi_{a_j})<\sum_{b\in H} u_{a_j}(\pi_b)$ holds for each $a_j\in G$. Finally, this conclusion leads to a contradiction with the $(p-1,q)$-group envy-freeness of $\pi$. The result follows. \myqed
\end{myproof}

By Examples~\ref{exp:one} and~\ref{exp:two}, the opposite direction of the implication in Theorem~\ref{thm:three} does not hold with 3 or more agents. With 2 agents, group-group envy-freeness is also equivalent to envy-freeness and proportionality. Finally, Theorem~\ref{thm:three} also holds for expected allocations.

\section{Group Pareto efficiency}\label{sec:gpe}

We continue with group Pareto efficiency properties for arithmetic-mean group utilities. Our second main result is to give a taxonomy of strict implications between group Pareto efficiency notions for groups of fixed sizes (i.e.\ GPE$_{k}$ for fixed $k\in[1,n)$). We present the taxonomy in Figure~\ref{fig:petax}.

\begin{figure}[h]
\begin{center}
\begin{tabular}{ccc}
GPE$_{k+1}$ & $\Rightarrow$ & GPE$_{k}$
\end{tabular}
\end{center}
\caption{A taxonomy of group Pareto efficiency properties for fixed $k\in[1,n)$.}
\label{fig:petax}
\end{figure}

Our taxonomy contains $n$ group Pareto efficient axiomatic properties. By definition, we observe that $1$-group Pareto efficiency is equivalent to Pareto efficiency, and $n$-group Pareto efficiency to utilitarian efficiency. In fact, we next prove that the $k$th layer of properties in our taxonomy is exactly between the $(k-1)$th and $(k+1)$th layers. It then follows that $k$-group Pareto efficiency implies $j$-group Pareto efficiency for any $k\in[1,n]$ and $j\in[1,k]$. We now show this result for actual allocations.

\begin{mytheorem}\label{thm:four}
For $k\in[1,n]$, $j\in[1,k]$ and arithmetic-mean group utilities, we have that \emph{GPE}$_{k}$ implies \emph{GPE}$_{j}$ holds.
\end{mytheorem}

\begin{myproof}
The proof is by backward induction on $h\in[j,k]$ for a given allocation $\pi$. For $h=k$, the proof is trivial. For $h>j$, suppose that $\pi$ is $h$-group Pareto efficient. For $h=j$, let us assume that $\pi$ is not $j$-group Pareto efficient. We write $G_j$ for the fact that group $G$ has $j$ agents. We derive that there is $\pi^{\prime}$ such that both inequalities (\ref{eq:agefeight}) and (\ref{eq:agefnine}) hold.

\begin{equation}\label{eq:agefeight}
\forall G_{j}:\sum_{a\in G_{j}} u_a(\pi^{\prime}_a)\geq \sum_{a\in G_{j}} u_a(\pi_a)
\end{equation} 
\begin{equation}\label{eq:agefnine}
\exists H_{j}:\sum_{b\in H_{j}} u_b(\pi^{\prime}_b)>\sum_{b\in H_{j}} u_b(\pi_b)
\end{equation} 

We next show that $\pi^{\prime}$ dominates $\pi$ in a $(j+1)$-group Pareto sense. That is, we show that inequalities (\ref{eq:ageften}) and (\ref{eq:agefeleven}) hold.

\begin{equation}\label{eq:ageften}
\forall G_{(j+1)}:\sum_{a\in G_{(j+1)}} u_a(\pi^{\prime}_a)\geq \sum_{a\in G_{(j+1)}} u_a(\pi_a)
\end{equation} 
\begin{equation}\label{eq:agefeleven}
\exists H_{(j+1)}:\sum_{b\in H_{(j+1)}} u_b(\pi^{\prime}_b)>\sum_{b\in H_{(j+1)}} u_b(\pi_b)
\end{equation} 

We start with inequality (\ref{eq:ageften}). Let $G_{(j+1)}$ be a group of $(j+1)$ agents for which inequality (\ref{eq:ageften}) does not hold. Further, let $G^a_j=G_{(j+1)}\setminus\lbrace a\rbrace$ be a group of $j$ agents obtained from $G_{(j+1)}$ by excluding agent $a\in G_{(j+1)}$. By the fact that inequality (\ref{eq:agefeight}) holds for $G^a_j$, we conclude that $u_a(\pi^{\prime}_a)<u_a(\pi_a)$ holds for each $a\in G_{(j+1)}$. We can now form a set of $j$ agents such that inequality (\ref{eq:agefeight}) is violated for $\pi^{\prime}$. Hence, inequality (\ref{eq:ageften}) must hold. We next show that inequality (\ref{eq:agefeleven}) holds as well. Let $H_{(j+1)}$ be an arbitrary group of $(j+1)$ agents for which inequality (\ref{eq:agefeleven}) does not hold. By inequality (\ref{eq:agefeight}), we derive $u_b(\pi^{\prime}_b)\leq u_b(\pi_b)$ for each $b\in H_{(j+1)}$. There cannot exist a group of $j$ agents for which inequality (\ref{eq:agefnine}) holds for $\pi^{\prime}$. Hence, inequality (\ref{eq:agefeleven}) must hold. Finally, as both inequalities (\ref{eq:ageften}) and (\ref{eq:agefeleven}) hold, $\pi$ is not $(j+1)$-group Pareto efficient. This is a contradiction.
\myqed
\end{myproof}

The implication in Theorem~\ref{thm:four} does not reverse. Indeed, an allocation that is $1$-group Pareto efficient might not be $k$-group Pareto efficient even for $k=2$ and 2 agents. We illustrate this in Example~\ref{exp:three}.

\begin{myexample}\label{exp:three}  
Let us consider the fair division of 2 items $o_1,o_2$ between 2 agents $a_1,a_2$. Further, suppose that $a_1$ likes $o_1$ with 1 and $o_2$ with 2, whilst $a_2$ likes $o_1$ with 2 and $o_2$ with 1. The allocation $\pi_1$ that gives both items to $a_1$ is $1$-group Pareto efficient (i.e.\ Pareto efficient) but not $2$-group Pareto efficient (i.e.\ utilitarian efficient). To see this, note that $\pi_1$ is $2$-group Pareto dominated by another allocation $\pi_2$ that gives $o_2$ to $a_1$ and $o_1$ to $a_2$. The utility of the group of two agents is $3/2$ in $\pi_1$ and $2$ in $\pi_2$.\myqed
\end{myexample}

We next consider expected allocations. We know that an expected allocation that is Pareto efficient can be represented as a convex combination over actual allocations that are Pareto efficient \cite{bogomolnaia2001} (cited by 502 other papers in Google Scholar). This result holds for actual allocations as well. We generalize this result to our setting with groups of agents and bundles of items. That is, we show that a $k$-group Pareto efficient expected allocation can be represented as a combination over $k$-group Pareto efficient actual allocations. We believe that our result is much more general than the existing one because it holds for arbitrary groups and bundle utilities (e.g.\ monotone, additive, modular, etc.). In contrast, not each convex combination over Pareto efficient actual allocations represents an expected allocation that is Pareto efficient \cite{bogomolnaia2001}. This observation holds in our setting as well. 

\begin{mytheorem}\label{thm:five}
For $k\in[1,n]$, a $k$-group Pareto efficient expected allocation can be represented as a convex combination over $k$-group Pareto efficient actual allocations.
\end{mytheorem}

\begin{myproof}
Let $\Pi_1$ denote an expected allocation that is $k$-group Pareto efficient and $c_1$ be a convex combination over group Pareto efficient allocations that represents $\Pi_1$. Further, let us assume that $\Pi_1$ cannot be represented as a convex combination over $k$-group Pareto efficient allocations. Therefore, there are two types of allocations in $c_1$: (1) allocations that are $j$-group Pareto efficient for some $j\geq k$ and (2) allocations that are $j$-group Pareto efficient ex post for some $j<k$. By Theorem~\ref{thm:four}, allocations of type (1) are $k$-group Pareto efficient. And, by assumption, allocations of type (2) are not $g$-group Pareto efficient for any $g>j$. Let us consider such an allocation $\pi$ in $c_1$ of type (2) that is not $k$-group Pareto efficient. Hence, $\pi$ can be $k$-group Pareto improved by some other allocation $\pi^{\prime}$. We can replace $\pi$ with $\pi^{\prime}$ in $c_1$ and thus construct a new convex combination $c_{1,\pi}$. We can repeat this for some other allocation in $c_{1,\pi}$ of type (2) that is not $k$-group Pareto efficient. We thus eventually can construct a convex combination $c_2$ over $k$-group Pareto efficient ex post allocations with the following properties: (1) there is an allocation $\pi_2$ in $c_2$ for each allocation $\pi_1$ in $c_1$ and (2) the weight of $\pi_2$ in $c_2$ is equal to the weight of $\pi_1$ in $c_1$. Let $\Pi_2$ denote the allocation represented by $c_2$. 

Let $c_1$ be over $\pi_1$ to $\pi_h$ such that $\pi_1$ to $\pi_i$ are $k$-group Pareto efficient and $\pi_{i+1}$ to $\pi_h$ are not group $k$-Pareto efficient. Further, by construction, let $c_2$ be over $\pi_1$ to $\pi_i$ and $\pi^{\prime}_{i+1}$ to $\pi^{\prime}_h$ such that $\pi^{\prime}_{g}$ $k$-group Pareto dominates $\pi_{g}$ for each $g\in[i+1,h]$. We derive $\sum_{a_l\in G}  (u_{a_l}(\pi^{\prime}_g)-u_{a_l}(\pi_g))\geq 0$ for each group $G$ of $k$ agents and $\sum_{a_l\in H}  (u_{a_l}(\pi^{\prime}_g)-u_{a_l}(\pi_g))>0$ for some group $H$ of $k$ agents. The expected utility $\overline{u}_{a_l}(\Pi_1)$ of agent $a_l$ in combination $c_1$ is equal to $\sum_{g\in[1,i]} w(\pi_g)\cdot u_{a_l}(\pi_g)+\sum_{g\in[i+1,h]} w(\pi_g)\cdot u_{a_l}(\pi_g)$. The expected utility $\overline{u}_{a_l}(\Pi_2)$ of agent $a_l$ in combination $c_2$ is equal to $\sum_{g\in[1,i]} w(\pi_g)\cdot u_{a_l}(\pi_g)+\sum_{g\in[i+1,h]} w(\pi_g)\cdot u_{a_l}(\pi^{\prime}_g)$. Therefore, $\sum_{a_l\in G}  (\overline{u}_{a_l}(\Pi_2)-\overline{u}_{a_l}(\Pi_1))\geq 0$ holds for each group $G$ of $k$ agents and $\sum_{a_l\in H}  (\overline{u}_{a_l}(\Pi_2)-\overline{u}_{a_l}(\Pi_1))>0$ holds for some group $H$ of $k$ agents. Hence, $\Pi_2$ $k$-group Pareto dominates $\Pi_1$. This is a contradiction with the $k$-group Pareto efficiency of $\Pi_1$.  
\myqed
\end{myproof}

Theorem~\ref{thm:five} suggests that there are fewer $k$-group Pareto efficient allocations than $j$-group Pareto efficient allocations for $j\in[1,k]$. In fact, there can be substantially fewer such allocations even with 2 agents. We illustrate this in Example~\ref{exp:four}.

\begin{myexample}\label{exp:four}
Let us consider again the instance in Example~\ref{exp:three}. Further, consider the expected allocation $\Pi_{\epsilon}$ in which agent $a_1$ receives item $o_1$ with probability 1 and item $o_2$ with probability $1-\epsilon$, and agent $a_2$ receives item $o_2$ with probability $\epsilon$. In $\Pi_{\epsilon}$, $a_1$ receives expected utility $3 - 2\epsilon$ and $a_2$ receives expected utility $\epsilon$. For each fixed $\epsilon\in[0,1/2)$, $\Pi_{\epsilon}$ is $1$-group Pareto efficient (i.e.\ Pareto efficient). Hence, there are infinitely many such allocations. By comparison, there is just one $2$-group Pareto efficient (i.e.\ utilitarian efficient) allocation that gives to each agent the item they like with $2$.\myqed 
\end{myexample}

Interestingly, for an $n$-group Pareto efficient expected allocation, we can show both directions in Theorem~\ref{thm:five}. By definition, such allocations maximize the utilitarian welfare. We, therefore, conclude that an expected allocation is $n$-group Pareto efficient iff it can be represented as a convex combination over actual allocations that maximize the utilitarian welfare. Finally, Theorem~\ref{thm:four} and Example~\ref{exp:three} also hold for expected allocations and Theorem~\ref{thm:five} and Example~\ref{exp:four} also hold (trivially) for actual allocations. 

\section{Near group fairness}\label{sec:ngf}

Near group fairness relaxes group fairness. Our near notions are inspired by $\alpha$-fairness proposed in \cite{borsuk1933}. Let $k\in[1,n]$, $h\in[1,n]$ and $\alpha\in[0,1]$. We start with near group envy-freeness (i.e.\ GEF$^{\alpha}_{k,h}$). For given $k$ and $h$, we can always find a sufficiently small value for $\alpha$ such that a given allocation satisfies GEF$^{\alpha}_{k,h}$. Consequently, for given $k$ and $h$, there is \emph{always} some $\alpha$ such that at least one allocation is GEF$^{\alpha}_{k,h}$. By comparison, for given $k$ and $h$, allocations that satisfy GEF$_{k,h}$ may \emph{not} exist. Therefore, for given $k$, $h$ and $\alpha$, allocations that satisfy GEF$^{\alpha}_{k,h}$ may also \emph{not} exist. For example, note that GEF$_{k,h}$ is equivalent to GEF$^{\alpha}_{k,h}$ for each $k$, $h$ and $\alpha=1$. Moreover, for given $k$, $h$ and $\alpha$, we have that GEF$_{k,h}$ implies GEF$^{\alpha}_{k,h}$ holds. However, there might be allocations that are near $(k,h)$-group envy-free with respect to $\alpha$ but not $(k,h)$-group envy-free. We illustrate this for actual allocations in Example~\ref{exp:five}.

\begin{myexample}\label{exp:five}
Let us consider again the instance in Example~\ref{exp:one} and the allocation $\pi$ that gives to each agent the item they like with $3/2$. Recall that $\pi$ is not $(1,1)$-group envy-free (i.e.\ envy-free). Each agent assigns in $\pi$ utility $2$ to one of the other agents and $1$ to the other one. For $\alpha=3/4$, they assign in $\pi$ reduced utilities $2\alpha$, $\alpha$ to these agents. We conclude that $\pi$ is near $(1,1)$-group envy-free wrt $\alpha$ (i.e.\ $3/4$-envy-free).
\myqed
\end{myexample}

For a given $\alpha$, we can show that Theorems~\ref{thm:one},~\ref{thm:two} and~\ref{thm:three} hold for the notions GEF$^{\alpha}_{k,h}$ with any $k$ and $h$. We can thus construct an $\alpha$-taxonomy of near group envy-freeness concepts for each fixed $\alpha$. Moreover, for $\alpha_1,\alpha_2\in[0,1]$ with $\alpha_2\geq\alpha_1$, we observe that an allocation satisfies an $\alpha_2$-property in the $\alpha_2$-taxonomy only if the allocation satisfies the corresponding $\alpha_1$-property in the corresponding $\alpha_1$-taxonomy. We further note that GEF$^{\alpha_2}_{k,h}$ implies GEF$^{\alpha_1}_{k,h}$. By Example~\ref{exp:five}, this implication does not reverse.

We proceed with near group Pareto efficiency (i.e.\ GPE$^{\alpha}_{k}$). For a given $k$, allocations satisfying GPE$_{k}$ \emph{always} exists. For given $k$ and $\alpha$, we immediately conclude that allocations satisfying GPE$^{\alpha}_{k}$ also \emph{always} exists. Similarly as for near group envy-freeness, GPE$_{k}$ is equivalent to GPE$^{\alpha}_{k}$ for each $k$ and $\alpha=1$, and GPE$_{k}$ implies GPE$^{\alpha}_{k}$ for each $k$ and $\alpha$. However, there might be allocations that are near $k$-group Pareto efficient with respect to $\alpha$ but not $k$-group Pareto efficient. We illustrate this for actual allocations in Example~\ref{exp:six}.

\begin{myexample}\label{exp:six}
Let us consider again the instance in Example~\ref{exp:three} and the allocation $\pi$ that gives to each agent the item they like with 1. This allocation is not $1$-group Pareto efficient (i.e.\ Pareto efficient) because each agent receives utility 2 if they swap items in $\pi$. For $\alpha=1/2$, $\pi$ is not $\alpha$-Pareto dominated by the allocation in which the items are swapped. Moreover, $\pi$ is not $\alpha$-Pareto dominated by any other allocation. We conclude that $\pi$ is near $1$-group Pareto efficient wrt $\alpha$ (i.e.\ $1/2$-Pareto efficient).
\myqed
\end{myexample}

For a given $\alpha$, we can also show that Theorem~\ref{thm:four} holds for the notions GPE$^{\alpha}_{k}$ with any $k$. We can thus construct an $\alpha$-taxonomy of near group Pareto efficiency properties for each fixed $\alpha$. In contrast to near group envy-freeness, allocations that satisfy an $\alpha$-property in an $\alpha$-taxonomy always exists. Also, for $\alpha_1,\alpha_2\in[0,1]$ with $\alpha_2\geq\alpha_1$, we observe that GPE$^{\alpha_2}_{k}$ implies GEF$^{\alpha_1}_{k}$ holds. By Example~\ref{exp:six}, we confirm that this is a strict implication. Theorem~\ref{thm:five} further holds for near $k$-group Pareto efficiency. Finally, Examples~\ref{exp:five} and~\ref{exp:six} hold for expected allocations as well.

\section{Prices of group fairness}\label{sec:pf}

We use \emph{prices of group fairness} and measure the ``loss'' in social welfare efficiency between different ``layers'' in our taxonomies. Our prices are inspired by the price of fairness proposed in \cite{bertsimas2011}. Prices of fairness are normally measured in the worst-case scenario. We proceed similarly and prove only the lower bounds of our prices for the \emph{u}tilitarian, the \emph{e}galitarian and the \emph{n}ash welfares in actual allocations.

\begin{mytheorem}\label{thm:six}
The prices $p^u_{\text{\tiny{GEF}}}$, $p^u_{\text{\tiny{GPE}}}$, $p^u_{\text{\tiny{FAIR}}}$ are all at least the number $n$ of agents, whereas the prices $p^e_{\text{\tiny{GEF}}}$, $p^e_{\text{\tiny{GPE}}}$, $p^e_{\text{\tiny{FAIR}}}$ and $p^n_{\text{\tiny{GEF}}}$, $p^n_{\text{\tiny{GPE}}}$, $p^n_{\text{\tiny{FAIR}}}$ are all unbounded.
\end{mytheorem}

\begin{myproof}
Let us consider the fair division of $n$ items to $n$ agents. Suppose that agent $a_i$ likes item $o_i$ with 1, and each other item with $\epsilon$ for some small $\epsilon\in(0,1)$. For $k\in[1,n]$, let $\pi_k$ denote an allocation in which $k$ agents receive items valued with 1 and $(n-k)$ agents receive items valued with $\epsilon$. By Theorem~\ref{thm:three}, $\pi_n$ is $k$-group envy-free as each agent receives their most valued item. By Theorem~\ref{thm:four}, $\pi_n$ is also $k$-group Pareto efficient. Further, for a fixed $k$, it is easy to check that $\pi_k$ is also $k$-group envy-free and $k$-group Pareto efficient. We start with the utilitarian prices. The utilitarian welfare in $\pi_n$ is $n$ whereas the one in $\pi_k$ is $k$ as $\epsilon$ goes to 0. Consequently, the corresponding ratios for ``layer'' $k$ in each taxonomy all go to $n/k$. Therefore, the corresponding prices go to $n$ as $k$ goes to $1$. We next give the egalitarian and Nash prices. The egalitarian and Nash welfares in $\pi_n$ are both equal to $1$. These welfares in $\pi_k$ are equal to $\epsilon$ and $\epsilon^{(n-k)}$ respectively. The corresponding ratios for ``layer'' $k$ in each taxonomy are then equal to $1/\epsilon$ and $1/\epsilon^{(n-k)}$. Consequently, the corresponding prices go to $\infty$ as $\epsilon$ goes to $0$. 
\myqed
\end{myproof}

Theorem~\ref{thm:six} holds for expected allocations as well. Finally, it also holds for near group fair allocations.

\section{Conclusions}\label{sec:con}

We studied the fair division of items to agents supposing agents can form groups. We thus proposed new group fairness axiomatic properties. Group envy-freeness requires that no group envies another group. Group Pareto efficiency requires that no group can be made better off without another group be made worse off. We analyzed the relations between these properties and several existing properties such as envy-freeness and proportionality. We generalized an important result from Pareto efficiency to group Pareto efficiency. We moreover considered near group fairness properties. We finally computed three prices of group fairness between such properties for three common social welfares: the utilitarian welfare, the egalitarian welfare and the Nash welfare.

In future, we will study more group aggregators. For example, our results hold for arithmetic-mean group utilities (i.e.\ Theorems~\ref{thm:one}-\ref{thm:six}). We can however also show them for \emph{geometric-mean}, \emph{minimum}, or \emph{maximum} group utilities (i.e.\ the root of the product over agents' utilities for the bundle, the minimum over agents' utilities for the bundle, the maximum over agents' utilities for the bundle). We will also study the relations of our group properties to other fairness properties for individual agents such as \emph{min-max fair share}, \emph{max-min fair share} and \emph{graph envy-freeness}. Finally, we submit that it is also worth adapting our group properties to other fair division settings as well \cite{aleksandrov2015ijcai}.

\bibliographystyle{splncs04}
\bibliography{groups}

\begin{thebibliography}{10}
\providecommand{\url}[1]{\texttt{#1}}
\providecommand{\urlprefix}{URL }
\providecommand{\doi}[1]{https://doi.org/#1}

\bibitem{aleksandrov2015ijcai}
Aleksandrov, M., Aziz, H., Gaspers, S., Walsh, T.: Online fair division:
  analysing a food bank problem. In: Proceedings of the Twenty-Fourth {IJCAI}
  2015, Buenos Aires, Argentina, July 25-31, 2015. pp. 2540--2546 (2015)

\bibitem{aleksandrov2017mcm}
Aleksandrov, M., Walsh, T.: Most competitive mechanisms in online fair
  division. In: {KI} 2017: Advances in Artificial Intelligence. pp. 44--57
  (2017)

\bibitem{aleksandrov2017ijcai}
Aleksandrov, M., Walsh, T.: Pure {Nash} equilibria in online fair division. In:
  Sierra, C. (ed.) Proceedings of the Twenty-Sixth {IJCAI} 2017, Melbourne,
  Australia. pp. 42--48 (2017)

\bibitem{aziz2018}
Aziz, H., Bouveret, S., Caragiannis, I., Giagkousi, I., Lang, J.: Knowledge,
  fairness, and social constraints. In: Proceedings of the Thirty-Second {AAAI}
  2018, New Orleans, Louisiana, USA, February 2-7, 2018. AAAI Press (2018)

\bibitem{aziz2015epera}
Aziz, H., Mackenzie, S., Xia, L., Ye, C.: Ex post efficiency of random
  assignments. In: Proceedings of the 2015 International {AAMAS} Conference,
  Istanbul, Turkey, May 4-8, 2015. pp. 1639--1640. IFAAMAS (2015)

\bibitem{aziz2017}
Aziz, H., Rauchecker, G., Schryen, G., Walsh, T.: Algorithms for max-min share
  fair allocation of indivisible chores. In: Proceedings of the Thirty-First
  {AAAI} 2017, San Francisco, California, {USA}, February 4-9, 2017. pp.
  335--341. {AAAI} Press (2017)

\bibitem{bertsimas2011}
Bertsimas, D., Farias, V.F., Trichakis, N.: The price of fairness. Operations
  Research  \textbf{59}(1),  17--31 (2011)

\bibitem{bliem2016}
Bliem, B., Bredereck, R., Niedermeier, R.: Complexity of efficient and
  envy-free resource allocation: Few agents, resources, or utility levels. In:
  Proceedings of the Twenty-Fifth {IJCAI} 2016, New York, NY, USA, 9-15 July
  2016. pp. 102--108 (2016)

\bibitem{bogomolnaia2001}
Bogomolnaia, A., Moulin, H.: A new solution to the random assignment problem.
  Journal of Economic Theory  \textbf{100}(2),  295--328 (2001)

\bibitem{bogomolnaia2016}
Bogomolnaia, A., Moulin, H., Sandomirskiy, F., Yanovskaya, E.: Dividing goods
  and bads under additive utilities. CoRR  \textbf{abs/1610.03745} (2016)

\bibitem{borsuk1933}
Borsuk, K.: Drei {S}ätze \"{u}ber die n-dimensionale euklidische {S}ph\"{a}re.
  Fundamenta Mathematicae  \textbf{20}(1),  177--190 (1933)

\bibitem{bouveret2017}
Bouveret, S., Cechl{\'{a}}rov{\'{a}}, K., Elkind, E., Igarashi, A., Peters, D.:
  Fair division of a graph. In: Proceedings of the Twenty-Sixth {IJCAI} 2017,
  August 19-25, 2017. pp. 135--141 (2017)

\bibitem{bouveret2008}
Bouveret, S., Lang, J.: Efficiency and envy-freeness in fair division of
  indivisible goods: Logical representation and complexity. Journal of AI
  Research {(JAIR)}  \textbf{32},  525--564 (2008)

\bibitem{brams2000}
Brams, S.J., Fishburn, P.C.: Fair division of indivisible items between two
  people with identical preferences: Envy-freeness, pareto-optimality, and
  equity. Social Choice and Welfare  \textbf{17}(2),  247--267 (Mar 2000)

\bibitem{brams2005}
Brams, S.J., King, D.L.: Efficient fair division: Help the worst off or avoid
  envy? Rationality and Society  \textbf{17}(4),  387--421 (2005)

\bibitem{brams1996}
Brams, S.J., Taylor, A.D.: Fair division - from cake-cutting to dispute
  resolution. Cambridge University Press (1996)

\bibitem{clippel2008}
de~Clippel, G.: Equity, envy and efficiency under asymmetric information.
  Economics Letters  \textbf{99}(2),  265--267 (2008)

\bibitem{pia2014}
Davidson, P., Evans, R.: Poverty in {Australia}. ACOSS (2014)

\bibitem{debreu1967}
Debreu, G.: Preference functions on measure spaces of economic agents.
  Econometrica  \textbf{35}(1),  111--122 (1967)

\bibitem{pia2016}
Dorsch, P., Phillips, J., Crowe, C.: Poverty in {Australia}. ACOSS (2016)

\bibitem{dubins1961}
Dubins, L.E., Spanier, E.H.: How to cut a cake fairly. The American
  Mathematical Monthly  \textbf{68}(1),  1--17 (1961)

\bibitem{hill1983}
Hill, T.P.: Determining a fair border. The American Mathematical Monthly
  \textbf{90}(7),  438--442 (1983)

\bibitem{husseinov2011}
Husseinov, F.: A theory of a heterogeneous divisible commodity exchange
  economy. Journal of Mathematical Economics  \textbf{47}(1),  54--59 (2011)

\bibitem{kaleta2014}
Kaleta, M.: Price of fairness on networked auctions. Journal of Applied
  Mathematics  \textbf{2014}, ~1--7 (2014)

\bibitem{keijzer2009}
de~Keijzer, B., Bouveret, S., Klos, T., Zhang, Y.: On the complexity of
  efficiency and envy-freeness in fair division of indivisible goods with
  additive preferences. In: Proceedings of the First International {ADT}
  Conference, Venice, Italy, October 20-23. pp. 98--110 (2009)

\bibitem{kokoye2013}
Kokoye, S.E.H., Tovignan, S.D., Yabi, J.A., Yegbemey, R.N.: Econometric
  modeling of farm household land allocation in the municipality of {B}anikoara
  in northern {B}enin. Land Use Policy  \textbf{34},  72--79 (2013)

\bibitem{lahaie2009}
Lahaie, S., Parkes, D.C.: Fair package assignment. In: Auctions, Market
  Mechanisms and Their Applications, First International {ICST} Conference,
  {AMMA} 2009, Boston, MA, USA, May 8-9, 2009, Revised Selected Papers. p.~92
  (2009)

\bibitem{lumet2012}
Lumet, C., Bouveret, S., Lema{\^{\i}}tre, M.: Fair division of indivisible
  goods under risk. In: {ECAI}. Frontiers in AI and Applications, vol.~242, pp.
  564--569. {IOS} Press (2012)

\bibitem{manurangsi2017}
Manurangsi, P., Suksompong, W.: Computing an approximately optimal agreeable
  set of items. In: Proceedings of the Twenty-Sixth {IJCAI} 2017, Melbourne,
  Australia, August 19-25, 2017. pp. 338--344 (2017)

\bibitem{nicosia2017}
Nicosia, G., Pacifici, A., Pferschy, U.: Price of fairness for allocating a
  bounded resource. European Journal of Operational Research  \textbf{257}(3),
  933--943 (2017)

\bibitem{parkes2015}
Parkes, D.C., Procaccia, A.D., Shah, N.: Beyond dominant resource fairness:
  Extensions, limitations, and indivisibilities. {ACM} Transactions
  \textbf{3}(1),  1--22 (2015)

\bibitem{schmeidler1972}
Schmeidler, D., Vind, K.: Fair net trades. Econometrica  \textbf{40}(4),
  637--642 (1972)

\bibitem{halevi2015}
Segal{-}Halevi, E., Nitzan, S.: Fair cake-cutting among groups. CoRR
  \textbf{abs/1510.03903} (2015)

\bibitem{halevi2018}
Segal{-}Halevi, E., Suksompong, W.: Democratic fair division of indivisible
  goods. In: Proceedings of the Twenty-Seventh {IJCAI-ECAI} 2018, Stockholm,
  Sweden, July 13-19, 2018 (2018), to appear

\bibitem{smet2016}
Smet, P.: Nurse rostering: models and algorithms for theory, practice and
  integration with other problems. 4OR  \textbf{14}(3),  327--328 (2016)

\bibitem{steinhaus1948}
Steinhaus, H.: The problem of fair division. Econometrica  \textbf{16}(1),
  101--104 (1948)

\bibitem{stone1942}
Stone, A.H., Tukey, J.W.: Generalized “sandwich” theorems. Duke
  Mathematical Journal  \textbf{9}(2),  356--359 (06 1942)

\bibitem{suksompong2016}
Suksompong, W.: Assigning a small agreeable set of indivisible items to
  multiple players. In: Proceedings of the Twenty-Fifth {IJCAI} 2016, New York,
  NY, USA, 9-15 July 2016. pp. 489--495. {IJCAI/AAAI} Press (2016)

\bibitem{suksompong2018}
Suksompong, W.: Approximate maximin shares for groups of agents. Mathematical
  Social Sciences  \textbf{92},  40--47 (2018)

\bibitem{todo2011}
Todo, T., Li, R., Hu, X., Mouri, T., Iwasaki, A., Yokoo, M.: Generalizing
  envy-freeness toward group of agents. In: Proceedings of the Twenty-Second
  {IJCAI} 2011, Barcelona, Catalonia, Spain, July 16-22, 2011. pp. 386--392
  (2011)

\bibitem{varian1974}
Varian, H.R.: Equity, envy, and efficiency. Journal of Economic Theory
  \textbf{9}(1),  63--91 (1974)

\bibitem{vind1964}
Vind, K.: Edgeworth-allocations in an exchange economy with many traders.
  International Economic Review  \textbf{5}(2),  165--177 (1964)

\bibitem{weller1985}
Weller, D.: Fair division of a measurable space. Journal of Mathematical
  Economics  \textbf{14}(1),  5 -- 17 (1985)

\bibitem{yokoo2003}
Yokoo, M.: Characterization of strategy/false-name proof combinatorial auction
  protocols: Price-oriented, rationing-free protocol. In: Proceedings of the
  Eighteenth {IJCAI} 2003, Acapulco, Mexico, August 9-15, 2003. pp. 733--742
  (2003)

\bibitem{zhou1992}
Zhou, L.: Strictly fair allocations in large exchange economies. Journal of
  Economic Theory  \textbf{57}(1),  158--175 (1992)

\end{thebibliography}

\end{document}